# Atomically Engineered $Hf_{0.5}Zr_{0.5}O_2$ Integrated Nano-Electromechanical Transducers

*Mayur Ghatge, Glen Walters, Toshikazu Nishida, Roozbeh Tabrizian*[*]

Electrical and Computer Engineering Department, University of Florida, Gainesville, Florida 32611, USA.

ABSTRACT: The monolithic integration of electromechanical transduction at the nanoscale with advanced CMOS is among the most important challenges of semiconductor electronic systems to leverage the multi-domain sensing, actuation, and resonance properties of nano-mechanical systems. Here we report on the demonstration of vibrating devices enabled by atomically engineered ferroelectric $Hf_{0.5}Zr_{0.5}O_2$ thin films with a variety of mechanical resonance modes with frequencies ($f_0$) between 340kHz – 13GHz and frequency-quality ($Q$) factor products ($f_0 \times Q$) up to $3.97 \times 10^{12}$. Experiments based on electrical and optical probing elucidate and quantify the role of the electrostrictive effect in the electromechanical transduction behavior of the $Hf_{0.5}Zr_{0.5}O_2$ film. We further demonstrate the role of nonlinear electromechanical scattering on the operation of $Hf_{0.5}Zr_{0.5}O_2$ transduced resonators. This investigation also highlights the potential of atomically engineered ferroelectric $Hf_{0.5}Zr_{0.5}O_2$ transducers for new classes of CMOS-monolithic linear and nonlinear nanomechanical resonators in centimeter- and millimeter-wave frequencies.

Integrated nano-electromechanical transducers enable extreme miniaturization of sensors and actuators to facilitate mechanical interaction with the outside world at the nano-scale with ultra-high resolution[1-3]. More importantly, integrated nano-electromechanical transducers can facilitate the harnessing of high-frequency and high-$Q$ mechanical resonance dynamics in semiconductor nano-structures to realize monolithically integrated stable frequency references and wideband spectral processors at centimeter- and millimeter-wave regimes[4-5]. Available frequency reference and filter technologies rely heavily on high-$Q$ micro-mechanical resonators implemented on semiconductor or insulator substrates. These resonators benefit from integrated capacitive gaps[6-7] or piezoelectric film transducers[8-9] that enable electrical excitation and sensing of high-$Q$ mechanical resonance modes over low-frequency (LF) to ultra-high-frequency (UHF) regimes. Extreme frequency scaling of mechanical resonators to operate over super- and extremely-high frequencies (i.e. SHF and EHF) requires radical miniaturization of their dimensions to few tens of nanometers. However, a proportional miniaturization of available integrated transducers is substantially limited by technological limitations for realization of ultra-narrow gaps and ultra-thin piezoelectric films. Further shrinking of capacitive gaps and conventional piezoelectric film transducers below 20nm drastically reduces their electromechanical transduction efficiency[10-11] and prevent them from detecting the vanishingly small vibration motions at the nanoscale beyond the thermomechanical noise level at the room temperature.

High-$Q$ nano-mechanical resonators based on two-dimensional (2D) crystals such as graphene and molybdenum disulfide have been extensively studied using non-integrated transduction schemes and demonstrated excellent potential for extreme frequency scaling through thickness miniaturization[12-17]. Besides the outstanding challenges with efficient integration of 2D crystals with CMOS, the absence of a compatible integrated nano-electromechanical transduction scheme prevents the harnessing of the promising frequency scalability and high-$Q$ properties for realization of monolithic frequency references and spectral processors.

Ultra-thin hafnium dioxide films have recently emerged as a new class of multi-morph atomic layered films that can be engineered to provide large ferroelectric properties[18-20]. These films have recently been studied for their potential to realize ultra-low-power and extremely miniaturized non-volatile memory devices. Hafnium dioxide nano-films can be deposited using a fully-CMOS compatible fabrication process to conformally cover the peripheral faces of semiconductor nano-structures. In this work, we demonstrate atomic layered ferroelectric hafnium zirconium oxide films ($Hf_{0.5}Zr_{0.5}O_2$) that are morphologically engineered by thermomechanical stress mediation to realize highly efficient integrated nano-electromechanical transducers based on the electrostrictive effect. $Hf_{0.5}Zr_{0.5}O_2$ layers as thin as 10nm are integrated on silicon (Si) and aluminum nitride (AlN) membranes to enable electromechanical transduction of a variety of mechanical resonance modes with frequencies ($f_0$) over 340kHz – 13GHz and frequency-quality ($Q$) factor products up to $3.97 \times 10^{12}$, surpassing the performance of 2D-crystal nano-mechanical resonators at similar thickness scales. Exploiting electrical and optical probing, we demonstrate and quantify the role of the nonlinear electrostrictive effect in the electromechanical transduction behavior of the $Hf_{0.5}Zr_{0.5}O_2$ film. We further demonstrate the contribution of nonlinear electromechanical scattering on the operation of $Hf_{0.5}Zr_{0.5}O_2$ transduced resonators which sheds light on the potential application and miniaturization roadmap for new classes of CMOS-monolithic linear and nonlinear nanomechanical resonators in centimeter- and millimeter-wave frequencies.

**RESULTS AND DISCUSSION**

**Atomic Engineering of $Hf_{0.5}Zr_{0.5}O_2$ and Resonator Processing.** Ultra-thin $Hf_{0.5}Zr_{0.5}O_2$ is known to have various stable and meta-stable crystal phases[21-22]. Among these, the non-centrosymmetric orthorhombic meta-stable phase provides unique ferroelectric properties[23-24]. While $Hf_{0.5}Zr_{0.5}O_2$ films grown by atomic layer deposition (ALD) are amorphous in nature due to the low thermal budget of the growth process, the use of a titanium nitride (TiN) capping layer and application of rapid thermal annealing (RTA) on the stack promotes the dominant growth of the ferroelectric orthorhombic texture (see Supporting Information, $S_1$). **Figure 1** demonstrates the cross-sectional transmission electron microscopy (XTEM) image of the TiN-on-$Hf_{0.5}Zr_{0.5}O_2$ stack (**Figure 1a**), the fast-Fourier transformation (FFT) of the atomic diffraction pattern highlighting the single-crystal orthorhombic phase after RTA (**Figure 1b**), and ferroelectric polarization-electric field characteristic (**Figure 1c**), directly measured after RTA and wake-up process. The TiN-on-$Hf_{0.5}Zr_{0.5}O_2$ stack is deposited on two substrates with (a) 70nm single-crystal silicon (Si) thickness and (b) 320nm stack of sputtered hexagonal aluminum nitride (AlN), molybdenum (Mo), and Si. The two substrates are used for fabrication of various mechanical resonator devices. We employed photolithography, reactive-ion and wet etching process, and deposition of platinum electrodes to realize free-standing membranes and facilitate

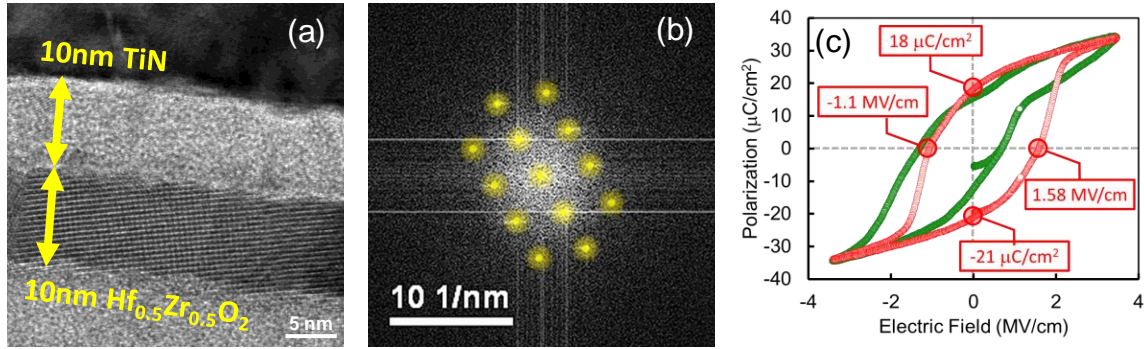

**Figure 1.** (a) Cross-sectional transmission electron microscopy (XTEM) image of TiN-on-$Hf_{0.5}Zr_{0.5}O_2$ stack showing the atomic diffraction pattern of the 10nm film; (b) FFT of the ultra-thin $Hf_{0.5}Zr_{0.5}O_2$ film showing single crystal orthorhombic texture; (c) measured polarization-electric field characteristic of the $Hf_{0.5}Zr_{0.5}O_2$ film after RTA (green curve) and subsequent wake-up process (red curve) showing the hysteresis behavior corresponding to the ferroelectric effect. electrical probing pads to access the 10nm $Hf_{0.5}Zr_{0.5}O_2$ integrated nano-electromechanical transducer (see Supporting Information, $S_1$, for fabrication processing details).

**Nanomechanical Resonators with 10nm $Hf_{0.5}Zr_{0.5}O_2$ Integrated Transducers.** We first demonstrate various resonators operating in flexural and extensional mechanical resonance modes over 340kHz to 13GHz frequency range. In these devices, the $Hf_{0.5}Zr_{0.5}O_2$ transducers are polarized through application of 3.5 volts DC for ~1s to provide linear electromechanical transduction (see Supporting Information, $S_2$, for further details on polling) upon insertion of ac signal. **Figure 2** shows the scanning electron microscope (SEM) and XTEM images of the $Hf_{0.5}Zr_{0.5}O_2$ transduced Si resonator. This device is excited through application of stroboscopic voltage signal across the $Hf_{0.5}Zr_{0.5}O_2$ film, and the flexural resonance vibration at 339kHz is detected using a holographic optical probe. A *Q* of 250 is extracted, while operating in air at room temperature.

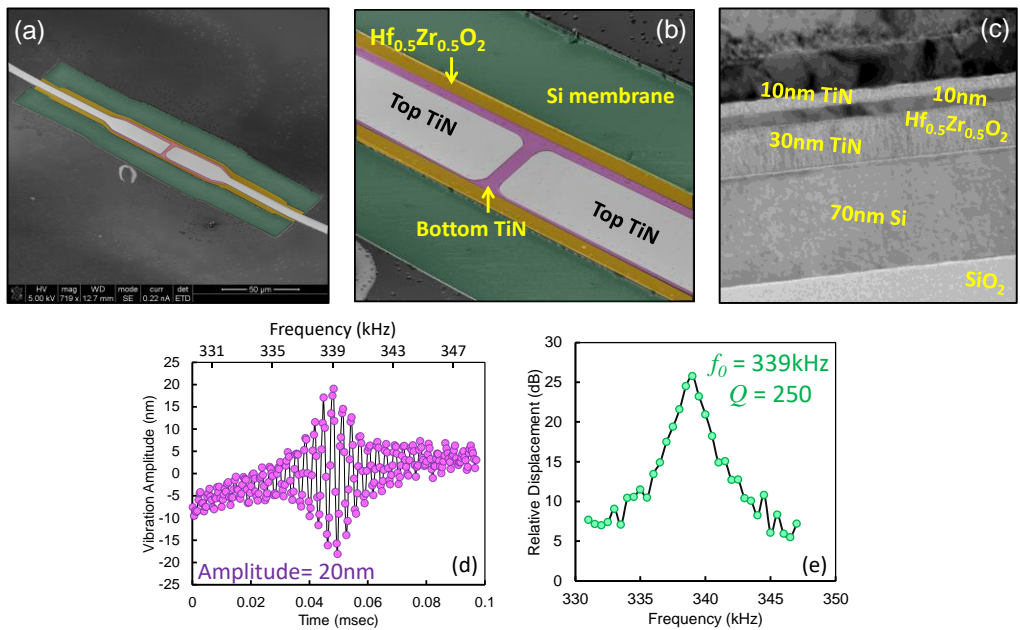

**Figure 2.** (a) SEM image of $Hf_{0.5}Zr_{0.5}O_2$ transduced Si nanomechanical resonator operating in out-of-plane flexural mode; (b) zoomed-in SEM of the resonator around TiN top electrodes used for application of stroboscopic ac voltage; (c) XTEM image of the resonator showing the thickness of transducer and Si layers; (d) stroboscopic vibration analysis using optical probe at the center of the resonator, showing a vibration amplitude of ~20nm at resonance; (e) the frequency response of the $Hf_{0.5}Zr_{0.5}O_2$ transduced Si resonator extracted from stroboscopic analysis showing the flexural resonance peak at 339kHZ and a $Q$ of 250.

**Figure 3** shows the SEM and XTEM images of two ultra- and super-high-frequency (UHF and SHF: 0.3-30GHz) $Hf_{0.5}Zr_{0.5}O_2$ transduced resonators implemented in AlN-on-Si platform, and their frequency response. The 120nm AlN piezoelectric transducer is used along with the 10nm $Hf_{0.5}Zr_{0.5}O_2$ to enable two-port electrical transmission characterization of the resonators operating in UHF and SHF regimes. **Figure 3a and 3c** shows resonators operating in lateral- and thickness-extensional modes, with XTEM shown in **Figure 3b**. In these devices, the AlN film is used for actuation and 10nm $Hf_{0.5}Zr_{0.5}O_2$ film is used for electrical detection of the mechanical strain at resonance. The resonator in **Figure 3a** operates in the 3$^{rd}$ width-extensional resonance mode at 413MHz with a *Q* of 530. The resonator shown in **Figure 3c** operates in the fundamental thickness-extensional mode, where the resonance frequency is defined by the thickness of the resonator stack. The resonator shows a *Q* of 310 at 12.8GHz frequency, yielding the large $f_0 \times Q$ figure of merit of $3.97 \times 10^{12}$.

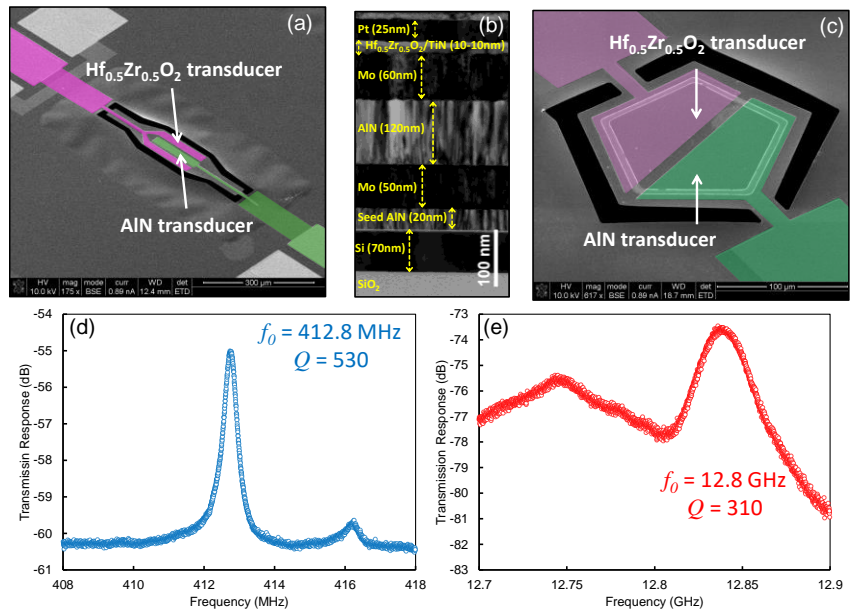

**Figure 3.** (a) SEM image of $Hf_{0.5}Zr_{0.5}O_2$ transduced AlN – on – Si resonator operating in in-plane extensional mode; (b) XTEM image of the $Hf_{0.5}Zr_{0.5}O_2$ transduced AlN – on – Si resonators showing the thickness of different layers in the stack; (c) SEM image of the $Hf_{0.5}Zr_{0.5}O_2$ transduced AlN – on – Si resonator operating in thickness – extensional mode; (d) the frequency response of the resonator in 3a, operating at 412.8 MHz with a *Q* of 530; (e) the frequency response of the resonator in 3c, operating in cm-wave frequency of 12.8 GHz, with a $f_0 \times Q$ product of $3.97 \times 10^{12}$.

**Electrostrictive Effect in Ferroelectric 10nm $Hf_{0.5}Zr_{0.5}O_2$ Transducers.** To gain insight and quantitative understanding of the electromechanical transduction physics in $Hf_{0.5}Zr_{0.5}O_2$ film, we perform experimental investigation and compare results with an analytical model. The electromechanical transduction in bulk ferroelectric materials is known to be based on the electrostrictive effect[25]. In this inherently nonlinear effect, the relation between polarization-induced mechanical strain ($\varepsilon_i$) and the dielectric spontaneous polarization ($P_j$) of the ferroelectric material is defined by:

$$\varepsilon_i = Q_{ij}\left(P_j(\vec{E})\right)^2 \quad (1).$$

Here, $Q_{ij}$ are the electrostrictive coefficients, and $P_j$ is related to the electric field vector ($\vec{E}$) through the characteristic polarization-electric field loop. To identify the contribution of nonlinear electrostrictive effect in atomically engineered $Hf_{0.5}Zr_{0.5}O_2$ transducers, we drive the resonator in **Figure 1** into mechanical resonance while changing a DC bias voltage ($V_{DC}$) along with the ac stroboscopic signal. **Figure 4 (left panel)** elucidates the variation of the frequency response of the $Hf_{0.5}Zr_{0.5}O_2$ transduced Si resonator, around the out-of-plane flexural resonance mode, with different $V_{DC}$ over -1.8V to 1.8V. Varying the DC bias changes the $Hf_{0.5}Zr_{0.5}O_2$ polarization, in accordance with the polarization-electric field loop shown in **Figure 1c** and results in a corresponding change in the vibration amplitude. **Figure 4 (right panel)** summarizes the vibration amplitude and phase for various $V_{DC}$ values. Specifically, **Figure 4c** compares the frequency response of the resonator when biased at 1.8V and −1.8V to elucidate the ~180° phase shift in vibration dynamics, when the polarization is reversed.

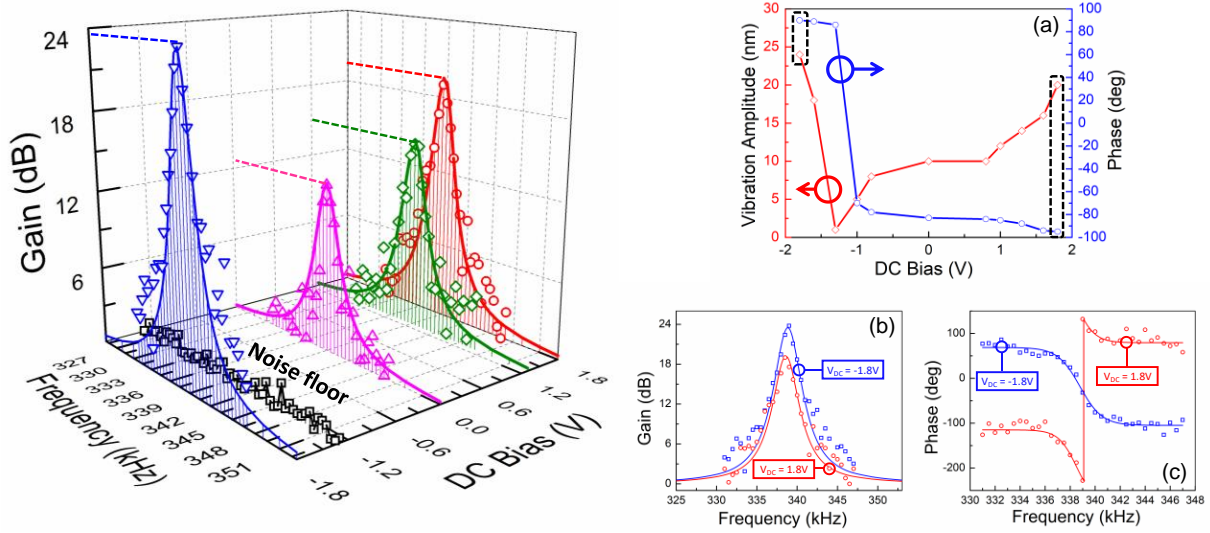

**Figure 4.** The frequency response of $Hf_{0.5}Zr_{0.5}O_2$ transduced Si resonator for different ferroelectric polarization scenarios. The left panel shows the frequency response, around the out-of-plane flexural resonance mode, for various DC bias voltages. On the right panel, (a) summarizes the resonance vibration amplitude and phase, for different DC bias voltages over −1.8V to 1.8V; (b) shows the vibration gain, relative to the noise floor, and (c) phase of the resonator when the $Hf_{0.5}Zr_{0.5}O_2$ transducer is polarized by 1.8 an −1.8 DC bias voltages. The ~180° difference in phase at resonance for the bias voltages of opposite sign elucidates the effect of polarization inversion on the electromechanical transduction.

**Nonlinear Electromechanical Scattering Effect on Resonator Performance.** To further understand the effect of nonlinear electrostrictive transduction on device operation, we extracted the magnitude of second and third order harmonics when the device is excited at the mechanical resonance frequency with different RF powers. The nonlinear harmonic generation in electromechanical resonators is attributed to various scattering mechanisms. These include resonator nonhomogeneous geometry, elastic anharmonicity of the constituent materials, thermal instabilities in elastic properties, and electromechanical scattering induced by the integrated transducer. While the thermomechanical scattering processes typically result in generation of third order harmonic, the electrostrictive effect in atomically engineered $Hf_{0.5}Zr_{0.5}O_2$ transducers scatters the electrical input into mechanical strain at twice the excitation frequency and yields second order harmonic generation. To quantify the effect of electromechanical scattering and compare it with thermomechanical nonlinearities, the harmonic generation dynamics in $Hf_{0.5}Zr_{0.5}O_2$ transduced AlN-on-Si resonator is studied. The resonator is excited using the $Hf_{0.5}Zr_{0.5}O_2$ transducer, with different electrical input powers over −12dBm to 18dBm, and the AlN piezoelectric transducer is used to sense mechanical vibrations at the output port. **Figure 5** shows the output power spectrum around the fundamental, second, and third harmonics for an $Hf_{0.5}Zr_{0.5}O_2$ transduced AlN-on-Si resonator that is excited in the lateral-extensional vibration mode at 255.5MHz.

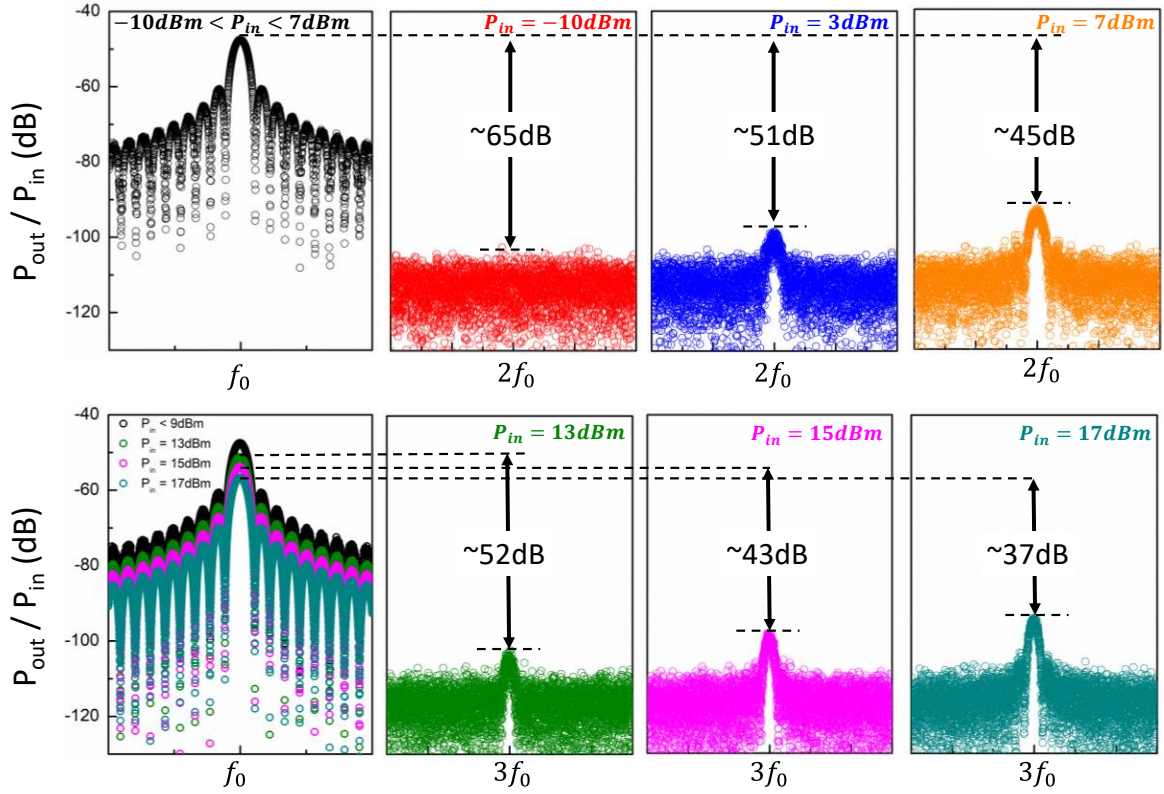

**Figure 5.** The output power spectrum around the fundamental, second and third harmonics for the $Hf_{0.5}Zr_{0.5}O_2$ transduced AlN-on-Si resonator operating in lateral-extensional mode, for various input RF powers. The top panel compares the magnitude of the second harmonic, created from the electromechanical scattering caused by the nonlinear electrostrictive effect. The bottom panel shows the relative magnitude of the third harmonic generated by the thermomechanical nonlinearities. The magnitude of the fundamental harmonic is reducing by an increase in power, when operating in mechanically nonlinear regime, which corresponds to the distortion of the linear frequency response of the resonator and increased insertion loss at higher input powers.

For input powers below 9dBm, the third order harmonic remains below the noise floor at −112dBm, confirming the operation of device in linear thermomechanical regime. However, the power of the second order harmonic increases with the increase in input RF power. As the input power exceeds 9dBm, the magnitude of the fundamental harmonic starts to decrease and the third harmonic stands out of the noise floor, with an increasing trend following the input power. This behavior corresponds to dominant contribution of thermomechanical nonlinearities that distort resonators linear transfer function and give rise to third harmonic generation[26-30]. **Figure 6** summarizes the second and third harmonic magnitudes over the wide range of input powers, highlighting different regimes of nonlinear scattering dominated by electromechanical and thermomechanical anharmonicities.

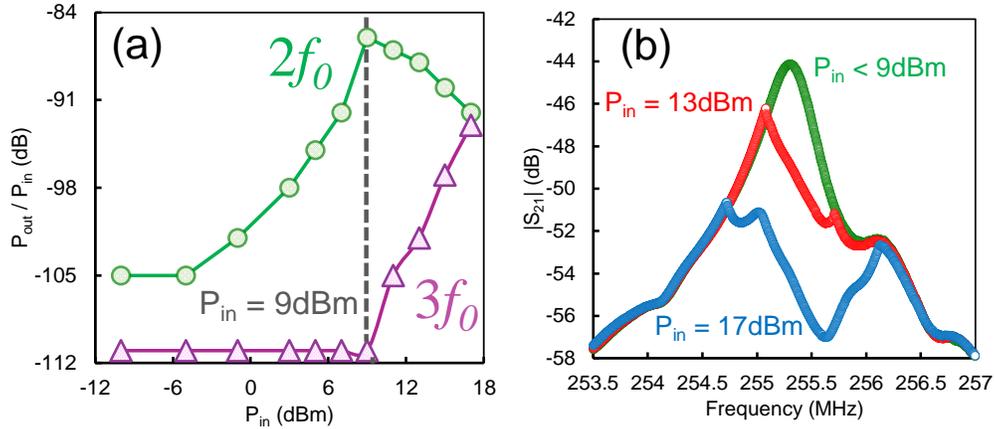

**Figure 6.** (a) The relative magnitude of second (circular markers in green) and third (triangle markers in purple) harmonics over different electrical drive powers, when the $Hf_{0.5}Zr_{0.5}O_2$ transduced AlN-on-Si resonator is excited in lateral-extensional mode. Below 9dBm electrical input power, the electromechanical scattering due to the nonlinear electrostrictive effect induce the 2$^{nd}$ harmonic with quadratically increasing magnitude. Beyond 9dBm, thermomechanical nonlinearities induce large third harmonics that stands out of the noise floor. The reduction in second harmonic magnitude beyond 9dBm input power is due to the distortion of resonator frequency response that increase the insertion loss at $f_0$. (b) The frequency response of the resonator when excited through $Hf_{0.5}Zr_{0.5}O_2$ transducer and detected through AlN piezoelectric transducer, for different input powers. Beyond the 9dBm input powers, the frequency response is distorted due to the rising thermomechanical nonlinearities.

For electrical excitations at the mechanical resonance with input powers below 9dBm, the resonator operates in linear mechanical regime. However, the electromechanical scattering induced by the electrostrictive effect in $Hf_{0.5}Zr_{0.5}O_2$ transducer results in excitation of second harmonic with quadratically proportional power. This result qualitatively follows the analytical derivations that quantify the relative ratio of linear and nonlinear mechanical strain excited with the application of electric field across the film thickness and through the electrostrictive effect (see Supporting Information, S$_2$) by

$$\frac{nonlinearly\ Scattered\ Mechanical\ Energy}{Linearly\ Transduced\ Mechanical\ Energy} \propto \left(\frac{\varepsilon_{nonlinear}}{\varepsilon_{linear}}\right)^2 = \frac{E_{ac}^2}{4P_0^2}\left(1 + \frac{\epsilon_3'}{\epsilon_3}E_{ac} + \frac{\epsilon_3'}{\epsilon_3}E_{ac}^2 + \cdots\right)^2 \quad (2).$$

Here, $\varepsilon_{linear}$ and $\varepsilon_{nonlinear}$ are linear and nonlinear strain excited by application of the electric filed $E_{ac}$ to the transducer; $P_0$ is the instantaneous polarization at zero electric field, and $\epsilon_3$ and $\epsilon_3'$ are the first- and second-order dielectric constants of the transducer film. Beyond 9dBm, the input electrical signal drives the resonator into thermomechanical nonlinearities, which gives rise to the third harmonic generation with a linearly increasing power. **Figure 6b** demonstrates the two-port transmission response of the resonator for various input powers, confirming the distortion in linear transfer function of the resonator due to the thermomechanical nonlinearities at input powers beyond 9dBm. It is also evident that the distortion in frequency response increases the insertion loss at the resonance frequency, which translates into an increase in device impedance at $f_0$ and induces a reduction in the relative power of electromechanically generated second harmonic, as evident in **Figure 6a**. Therefore, the experiment confirms the contribution of electrostrictive effect in nonlinear scattering and harmonic generation in Hf$_{0.5}$Zr$_{0.5}$O$_2$ transduced resonators. While the nonlinear electromechanical scattering resulted by electrostrictive effect in atomically engineered Hf$_{0.5}$Zr$_{0.5}$O$_2$ transducers degrades the resonator performance in mechanically linear regime, it provides a promising potential for nonlinear mode-coupling in nanosystems[31-33] or realization of novel nonlinear nano-acoustic and non-reciprocal components[34-35], especially at cm- and mm-wave frequencies.

**CONCLUSIONS**

In conclusion, we have demonstrated a new type of integrated nano-electromechanical transducer realized from 10nm ferroelectric $Hf_{0.5}Zr_{0.5}O_2$ films. The transducers were integrated on single crystal Si and AlN-on-Si membranes to realize high-$Q$ nanomechanical resonators over the 340kHZ to 13GHz frequency range, operating in flexural and extensional vibration modes. Among these, the 13GHz device demonstrated a figure of merit of $f_0 \times Q \approx 3.97 \times 10^{12}$ which is the highest reported for any nano-mechanical resonator operating in the cm-wave regime. Our study elucidates the role of the electrostrictive effect in the electromechanical transduction of $Hf_{0.5}Zr_{0.5}O_2$ transducers which enables precise control of resonator vibration amplitude and phase through variation of the ferroelectric polarization through the DC bias voltage. Also, we have demonstrated the contribution of the nonlinear electrostrictive effect on the electromechanical scattering of the resonator energy into the second harmonic signal. We have further compared the magnitude of the electromechanical scattering with the thermomechanical counterpart through monitoring the third harmonic excitation and linear frequency-response distortion over a wide electric input power range. This study confirms the large magnitude of the electromechanical scattering for $Hf_{0.5}Zr_{0.5}O_2$ transduced resonators operating in the linear thermomechanical regime and highlight an opportunity for realization of nonlinearly coupled nanomechanical systems for extreme frequency scaling and nonlinear coupling of frequency references, realization of wideband frequency combs, or implementation of integrated nonreciprocal spectral processors.

ASSOCIATED CONTENTS

**METHODS**

**Device Fabrication**. The $Hf_{0.5}Zr_{0.5}O_2$ nano-electromechanical transducers are realized through atomic layer deposition, stacking with TiN layer, and proper RTA. The nanomechanical resonators are fabricated by integration of $Hf_{0.5}Zr_{0.5}O_2$ films on Si and AlN-on-Si membranes, followed by various etching and deposition processes to create free-standing devices and enable electrical pads for application of excitation and sensing signals. (see Supporting Information, $S_1$, for details).

**Device Characterization.** $Hf_{0.5}Zr_{0.5}O_2$ films are electrically measured with a modified Sawyer-Tower circuit to extract polarization-electric field hysteresis characteristic corresponding to ferroelectric behavior. Nanomechanical resonators with integrated $Hf_{0.5}Zr_{0.5}O_2$ transducers are electrically characterized using Keysight N5222A network analyzer, Keysight E5173B signal generator and Keysight N9010A signal analyzer, to extract frequency response and characterize the magnitude of nonlinearly generated harmonics. The $Hf_{0.5}Zr_{0.5}O_2$ transduced Si nanomechanical resonator is optically characterized using a Lyncée tec R-2100 series reflection digital holographic microscope that is synchronized with a stroboscopic signal generator (see Supporting Information, $S_1$, for details).

**Scanning Electron Microscopy (SEM).** SEM images are taken using an FEI Nova NanoSEM 430 system.

**Transmission Electron Microscopy (TEM).** TEM images are taken using an FEI TECNAI F20 S/TEM system.

**Electrostrictive Effect Formulation.** The electromechanical transduction based on electrostrictive effect in ferroelectric materials can be formulated through polynomial approximation of the polarization-electric field characteristic. Two regimes of operation, i.e.

linear and nonlinear, can be identified depending on the electric field excitation amplitude and applied DC bias voltage (see Supporting Information, $S_2$, for details).

AUTHOR INFORMATION

**Corresponding Author**

*(rtabrizian@ece.ufl.edu).

**Author Contributions**

The manuscript was written through contributions of all authors. All authors have given approval to the final version of the manuscript. All authors contributed equally.

**Funding Sources**

This work was supported in part by the NSF grants ECCS 1610387 and ECCS 1752206.

REFERENCES

1. LaHaye, M. D., Buu, O., Camarota, B., & Schwab, K. C. (2004). Approaching the quantum limit of a nanomechanical resonator. *Science*, *304*(5667), 74-77.

2. Jensen, K., Kim, K., & Zettl, A. (2008). An atomic-resolution nanomechanical mass sensor. *Nature nanotechnology*, *3*(9), 533.

3. Barson, M. S., Peddibhotla, P., Ovartchaiyapong, P., Ganesan, K., Taylor, R. L., Gebert, M., ... & McCallum, J. (2017). Nanomechanical sensing using spins in diamond. *Nano letters*, *17*(3), 1496-1503.

4. Ramezany, A., & Pourkamali, S. (2018). Ultrahigh Frequency Nanomechanical Piezoresistive Amplifiers for Direct Channel-Selective Receiver Front-Ends. *Nano letters*, *18*(4), 2551-2556.

5. Villanueva, L. G., Karabalin, R. B., Matheny, M. H., Kenig, E., Cross, M. C., & Roukes, M. L. (2011). A nanoscale parametric feedback oscillator. *Nano letters*, *11*(11), 5054-5059.

6. Naing, T. L., Rocheleau, T. O., Ren, Z., Li, S. S., & Nguyen, C. T. C. (2016). High-Q UHF Spoke-Supported Ring Resonators. *Journal of Microelectromechanical Systems*, *25*(1), 11-29.

7. Pourkamali, S., Ho, G. K., & Ayazi, F. (2007). Low-impedance VHF and UHF capacitive silicon bulk acoustic wave resonators—Part I: Concept and fabrication. *IEEE Transactions on Electron Devices*, *54*(8), 2017-2023.

8. Ruby, R., Small, M., Bi, F., Lee, D., Callaghan, L., Parker, R., & Ortiz, S. (2012). Positioning FBAR technology in the frequency and timing domain. *IEEE transactions on ultrasonics, ferroelectrics, and frequency control*, *59*(3), 334-345.

9. Cassella, C., Oliva, N., Soon, J., Srinivas, M., Singh, N., & Piazza, G. (2017). Super High Frequency Aluminum Nitride Two-Dimensional-Mode Resonators With $k_t^2$ Exceeding 4.9%. *IEEE Microwave and Wireless Components Letters*, *27*(2), 105-107.

10. Iborra, E., Clement, M., Capilla, J., Olivares, J., & Felmetsger, V. (2012). Low-thickness high-quality aluminum nitride films for super high frequency solidly mounted resonators. *Thin Solid Films*, *520*(7), 3060-3063.

11. Sinha, N., Wabiszewski, G. E., Mahameed, R., Felmetsger, V. V., Tanner, S. M., Carpick, R. W., & Piazza, G. (2009, June). Ultra thin AlN piezoelectric nano-actuators. In *TRANSDUCERS 2009-2009 International Solid-State Sensors, Actuators and Microsystems Conference* (pp. 469-472). IEEE. Wang, Zenghui, Hao Jia, Xuqian Zheng, Rui Yang, Zefang Wang, G. J. Ye, X. H.

Chen, Jie Shan, and Philip X-L. Feng. "Black phosphorus nanoelectromechanical resonators vibrating at very high frequencies." *Nanoscale* 7, no. 3 (2015): 877-884.

12. Wang, Z., Jia, H., Zheng, X., Yang, R., Wang, Z., Ye, G. J., ... & Feng, P. X. L. (2015). Black phosphorus nanoelectromechanical resonators vibrating at very high frequencies. *Nanoscale*, *7*(3), 877-884.

13. Lee, J., Wang, Z., He, K., Shan, J., & Feng, P. X. L. (2013). High frequency MoS2 nanomechanical resonators. *ACS nano*, *7*(7), 6086-6091.

14. Bunch, J. S.; van der Zande, A. M.; Verbridge, S. S.; Frank, I. M.; Tanenbaum, D. M.; Parpia, J. M.; Craighead, H. G.; McEuen, P. L. Electromechanical Resonators from Graphene Sheets. Science 2007, 315, 490–493.

15. Chen, C. Y.; Rosenblatt, S.; Bolotin, K. I.; Kalb, W.; Kim, P.; Kymissis, I.; Stormer, H. L.; Heinz, T. F.; Hone, J. Performance of Monolayer Graphene Nanomechanical Resonators with Electrical Readout. Nat. Nanotechnol. 2009, 4, 861–867.

16. van der Zande, A. M.; Barton, R. A.; Alden, J. S.; Ruiz-Vargas, C. S.; Whitney, W. S.; Pham, P. H. Q.; Park, J. W.; Parpia, J. M.; Craighead, H. G.; McEuen, P. L. Large-Scale Arrays of Single Layer Graphene Resonators. Nano Lett. 2010, 10, 4869–4873.

17. Eichler, A.; Moser, J.; Chaste, J.; Zdrojek, M.; Wilson-Rae, I.; Bachtold, A. Nonlinear Damping in Mechanical Resonators Made from Carbon Nanotubes and Graphene. Nat. Nanotechnol. 2011, 6, 339–342.

18. Böscke, T. S., Müller, J., Bräuhaus, D., Schröder, U., & Böttger, U. (2011). Ferroelectricity in hafnium oxide thin films. *Applied Physics Letters*, *99*(10), 102903.

19. Mulaosmanovic, H., Ocker, J., Müller, S., Schroeder, U., Müller, J., Polakowski, P., ... & Slesazeck, S. (2017). Switching kinetics in nanoscale hafnium oxide based ferroelectric field-effect transistors. *ACS applied materials & interfaces*, *9*(4), 3792-3798.

20. Chernikova, A., Kozodaev, M., Markeev, A., Negrov, D., Spiridonov, M., Zarubin, S., ... & Gruverman, A. (2016). Ultrathin Hf0. 5Zr0. 5O2 ferroelectric films on Si. *ACS applied materials & interfaces*, *8*(11), 7232-7237.

21. Böscke, T. S., Teichert, S., Bräuhaus, D., Müller, J., Schröder, U., Böttger, U., & Mikolajick, T. (2011). Phase transitions in ferroelectric silicon doped hafnium oxide. *Applied Physics Letters*, *99*(11), 112904.

22. Hyuk Park, M., Joon Kim, H., Jin Kim, Y., Lee, W., Moon, T., & Seong Hwang, C. (2013). Evolution of phases and ferroelectric properties of thin Hf0. 5Zr0. 5O2 films according to the thickness and annealing temperature. *Applied Physics Letters*, *102*(24), 242905.

23. Materlik, R., Künneth, C., & Kersch, A. (2015). The origin of ferroelectricity in Hf1−xZrxO2: A computational investigation and a surface energy model. *Journal of Applied Physics*, *117*(13), 134109.

24. Schroeder, U., Yurchuk, E., Müller, J., Martin, D., Schenk, T., Polakowski, P., ... & Mikolajick, T. (2014). Impact of different dopants on the switching properties of ferroelectric hafniumoxide. *Japanese Journal of Applied Physics*, *53*(8S1), 08LE02.

25. Sundar, V., & Newnham, R. E. (1992). Electrostriction and polarization. *Ferroelectrics*, *135*(1), 431-446.

26. Gieseler, J., Novotny, L., & Quidant, R. (2013). Thermal nonlinearities in a nanomechanical oscillator. *Nature physics*, *9*(12), 806.

27. Segovia-Fernandez, J., & Piazza, G. (2013). Thermal nonlinearities in contour mode AlN resonators. *Journal of Microelectromechanical Systems*, *22*(4), 976-985.

28. Ghatge, M., Karri, P., & Tabrizian, R. (2017, January). Power-insensitive silicon crystal-cut for amplitude-stable frequency synthesis. In *2017 IEEE 30th International Conference on Micro Electro Mechanical Systems (MEMS)* (pp. 76-79). IEEE.

29. Polunin, P. M., Yang, Y., Dykman, M. I., Kenny, T. W., & Shaw, S. W. (2016). Characterization of MEMS resonator nonlinearities using the ringdown response. *Journal of Microelectromechanical Systems*, *25*(2), 297-303. Rocas, E., Collado, C., Mateu, J., Campanella, H., & O'Callaghan, J. M. (2008, June).

30. Third order intermodulation distortion in film bulk acoustic resonators at resonance and antiresonance. In *2008 IEEE MTT-S International Microwave Symposium Digest* (pp. 1259-1262). IEEE.

31. Matheny, M. H., Villanueva, L. G., Karabalin, R. B., Sader, J. E., & Roukes, M. L. (2013). Nonlinear mode-coupling in nanomechanical systems. *Nano letters*, *13*(4), 1622-1626.

32. Ghatge, M., & Tabrizian, R. (2018). Exploiting elastic anharmonicity in aluminum nitride matrix for phase-synchronous frequency reference generation. *Applied Physics Letters*, *112*(12), 123503.

33. Zhu, J., Ru, C. Q., & Mioduchowski, A. (2010). High-order subharmonic parametric resonance of multiple nonlinearly coupled micromechanical nonlinear oscillators. *Acta Mechanica*, *212*(1-2), 69-81.

34. Popa, B. I., & Cummer, S. A. (2014). Non-reciprocal and highly nonlinear active acoustic metamaterials. *Nature communications*, *5*, 3398.

35. Lepri, S., & Pikovsky, A. (2014). Nonreciprocal wave scattering on nonlinear string-coupled oscillators. *Chaos: An Interdisciplinary Journal of Nonlinear Science*, *24*(4), 043119.